\documentclass[review]{elsarticle}
\usepackage{url}

\usepackage{lineno,hyperref}
\usepackage{graphicx}%
\usepackage{multirow}%
\usepackage{amsmath,amssymb,amsfonts}%
\usepackage{amsthm}%
\usepackage{mathrsfs}%
\usepackage[title]{appendix}%
\usepackage{xcolor}%
\usepackage{textcomp}%
\usepackage{manyfoot}%
\usepackage{booktabs}%
\usepackage{algorithm}%
\usepackage{algorithmicx}%
\usepackage{algpseudocode}%
\usepackage{listings}%
\usepackage[export]{adjustbox}
\usepackage{makecell}
\modulolinenumbers[5]
\linenumbers

\journal{Journal of Magnetism and Magnetic Materials}

\begin{document}

\begin{frontmatter}

\title{Machine Learning Approach to Predict Curie Temperature in Binary Alloys\tnoteref{mytitlenote}}

\author{Svitlana Ponomarova\fnref{myfootnote}}
\author{Oleksandr Ponomarov\fnref{myfootnote}}
\author{Yurii Koval\fnref{myfootnote}}
\address{The G.V. Kurdyumov Institute for Metal Physics of the National Academy of Sciences of Ukraine}
\address{Akademika Vernadskoho Blvd., 36, Kyiv, Ukraine, 02000}
\author[mymainaddress]{sveta.ponomaryova@gmail.com}

\begin{abstract}
This study presents a machine learning approach to predict the Curie temperature in binary alloys, specifically focusing on the Fe-Pt, Fe-Ni, Fe-Pd, and Co-Pt compounds within a concentration range of 10 to 90 atomic percent. 

The optimal mathematical algorithm for this task is the Voting Ensemble algorithm, which combines the predictions from multiple individual models to produce a final prediction. The results are validated against classical methods for calculating Curie temperatures.

The experimental findings indicate that factors such as external pressure, atomic ordering, and alloy composition have a significant influence on the Curie temperatures in all examined binary systems. 
These factors can be leveraged to design alloys with specific Curie temperatures.  Moreover, the proposed features, feature analysis algorithms, and computational methods pave the way for advancements across various materials, including ternary alloys, bulk materials, and nanomaterials, inspiring innovation in the field.
\end{abstract}

\begin{keyword}
\texttt Machine learning \sep Curie temperature \sep Regression algorithms \sep Microsoft Azure
\end{keyword}

\end{frontmatter}

\linenumbers

\section{Introduction}

The Curie temperature is an important parameter in the study of magnetic materials. It indicates the temperature at which a material loses its magnetic order and transitions to a non-magnetic state.
The investigation of magnetic properties s typically carried out using experimental methods such as standard magnetometry, SQUID magnetometry \cite{Lewis}, inductively heated dilatometry \cite{Verma}, the magnetic circuit method \cite{Iorga}, among others.
There are several methods available for calculating the Curie temperature. Among these, Monte Carlo modeling is recognized as an effective approach, particularly for two-dimensional materials. Other methods include the mean field approximation \cite{Rusz} and calculations based on the Heisenberg Hamiltonian \cite{Ponomarova}.

However, traditional materials design methods often encounter challenges, such as lengthy development cycles and high costs.  Cloud technologies,
machine learning (ML), and artificial intelligence (AI) enable the design of
materials more quickly and cost-effectively, using traditional experiments primarily to validate ML results.
These modern approaches have been successfully applied in various fields, including metallic glasses, high-entropy alloys, shape-memory alloys, magnets, superalloys, catalysts, and structural materials.

Areas of materials research utilizing machine learning include:
\begin{itemize}
    \item autonomous material search using ML (examples for Fe-Pt-based alloys \cite{Iwasaki}, Ni-Pd alloys \cite{Iwasaki2})
    \item properties optimization performed for Fe-Pd  \cite{Gao2}
    \item accelerated materials design  successfully applied in Fe-based alloys \cite{Takahara}
    \item materials screening \cite{Hu}, \cite{Roy}, including screening for defects \cite{Berger}
    \item discovery of material properties, including low-dimensional systems \cite{Fang}.
\end{itemize}

Several research papers \cite{Jung}, \cite{Singh}, \cite{Brannvall}, \cite{Jung2}, \cite{Belot}, \cite{Hilgers}, \cite{Yang} focus on predicting the Curie temperature in different compounds. These studies contribute to the autonomous search for materials with specific properties \cite{Yuma} and to predicting the magnetic properties of materials \cite{Nyamnjoh}, particularly for spintronics applications \cite{Ho}. Furthermore, the integration of classical Density Functional Theory (DFT) and machine learning shows promise in predicting the Curie temperature of different materials \cite{Wang}.

Maintaining the properties of magnetic materials under operational conditions is essential for their effectiveness in practical applications. This research employs machine learning methods to investigate the factors that influence the Curie temperature in binary alloys of iron-group metals. In other words, it focuses on discovering material properties and designing new materials.
The Curie temperature can be tuned by several factors, including alloy composition and external pressure. Additionally, microstructural characteristics and the influence of other phase transitions, such as atomic ordering, play significant roles. In the case of the nanoscale, nanoparticle shape and size become predominant factors. Experimental investigations under these influences are shown in \cite{Rohman}, \cite{Mizoguchi}, \cite{Oomi}, and many other works.

The research focuses on Fe, Pt, Pd, Ni, and Co, which are combined into binary alloys: Fe-Pt, Fe-Pd, Fe-Ni, and Co-Pt.
The Fe-Pt alloys exhibit high magnetocrystalline anisotropy, making them good for high-performance permanent magnets \cite{Burkert}, \cite{Gehanno}. 
In Fe-Pd alloys, two ranges of alloy compositions are of particular interest. The first is the $L1_0$ ordered equiatomic Fe-Pd alloys, considered for ultrahigh-density magnetic recording media \cite{Wang}, \cite{Gehanno}. The second range, around 30 at.\% Pd, has been intensively studied due to its shape memory effect \cite{Cui}.
The Co-Pt alloys show ferromagnetic properties characterized by high coercivity and remanence, arising from their high anisotropy \cite{Islam}. These alloys can also undergo atomic ordering \cite{Pedan}, which alters their magnetic properties \cite{Tamion}.
The Fe-Ni alloys demonstrate high magnetic anisotropy, which is significantly influenced by the atomic ordering and composition of the alloy \cite{Qiao}, \cite{Nguyen}. 
Among the factors influencing Curie temperature is composition, which has been experimentally investigated for individual alloys, including Fe-Ni \cite{Liu}, Fe-Pt \cite{Vlaic}, and Fe-Pd \cite{Vlaic2}. Other important parameters are atomic ordering, particularly in alloys where this phenomenon can occur, and external pressure. Several experimental works have addressed these topics \cite{Wang2}, \cite{Oomi}, \cite{Ruban}, \cite{Aladerah}. 

The paper is organized as follows: Section 2 outlines the essential computational techniques, feature and dataset analysis, and training algorithms. Section 3.1 describes the training dataset, the optimal algorithm, feature importance analysis, and computational details specific to predicting Curie temperature. Section 3.2 presents the computational findings, highlighting the effects of external pressure, atomic ordering, and alloy composition on the Curie temperature. It also includes a thermodynamic analysis and a discussion of the key factors that influence it.

\section{Computational technique}
The regression method relies on a set of 'features' — the parameters of the model being studied — that determine the final result, known as 'labels.' In other words, 'features' provide the input data, while 'labels' represent the output.
We explore the regression in ML using different computational algorithms listed in the "Supplementary Materials" section.

Current research uses Microsoft Azure Machine Learning Studio to train, deploy, manage, and track machine learning models. Its advantages include the ability to scale computing resources up or down based on demand and integration with popular frameworks as TensorFlow, PyTorch, and Scikit-learn. 
 Azure's infrastructure is optimized for AI workloads for faster training and inference of ML models. 
 
 A good alternatives for Microsoft Azure could be Databricks ML \footnote{https://docs.databricks.com/en/machine-learning/train-model/}, TensorFlow \footnote{https://www.tensorflow.org/learn}, PyTorch \footnote{https://pytorch.org/docs/stable/}, Scikit-Learn \footnote{https://scikit-learn.org/stable/}.

The first steps in training the machine learning model are feature selection and dataset preparation. We tested several feature selection algorithms:
\newline
- MaxAbsScaler \footnote{https://scikit-learn.org/stable/modules/generated/sklearn.preprocessing.MaxAbsScaler.html}: \textit{transforms data dividing each feature by its maximum absolute value, to have the resulting data within the range of -1 to 1;}
\newline
- Sparse Normalizer \cite{Zhao}: \textit{selects a subset of essential features from a larger set;}
\newline
- Standard Scaler Wrapper \footnote{https://scikit-learn.org/stable/modules/generated/sklearn.preprocessing.StandardScaler.html}: \textit{standardizes features by removing the mean and scaling to unit variance;}
\newline
- TruncatedSVD Wrapper \cite{Falini}: \textit{reduces the number of features preserving the most essential information in the data. }

The data preparation phase also includes an analysis of dataset quality. To identify potential issues, the data guardrails procedure is conducted \footnote{https://learn.microsoft.com/en-us/azure/machine-learning/how-to-configure-auto-features?view=azureml-api-1}. 
One aspect is "missing feature values imputation," which substitutes missing values with alternatives to eliminate gaps in the input dataset. Additionally, "high cardinality" refers to datasets with a high level of detail and variation. In this case, the data pre-processing aims to reduce the number of unique features and lower dimensionality.

\begin{figure}[H]%
  \includegraphics[width=\textwidth, inner]{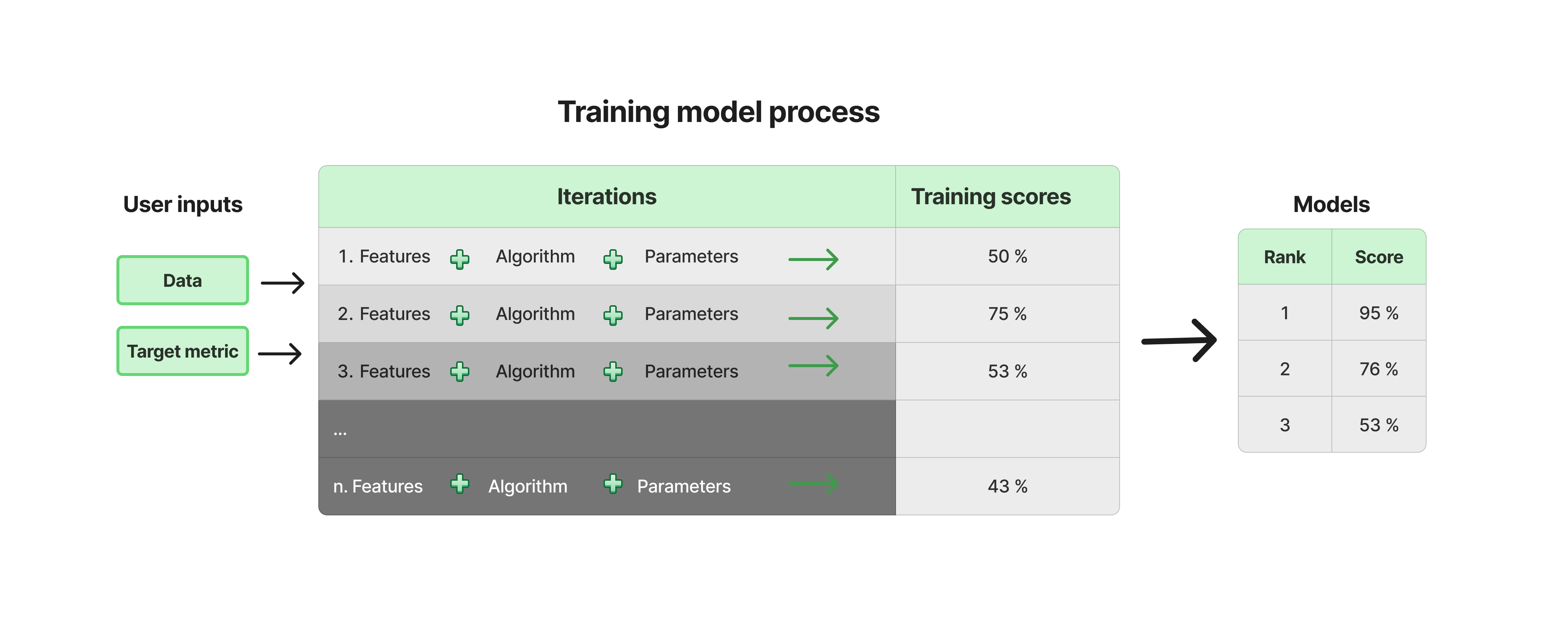}
  \caption{Workflow for training an ML model}
  \label{fig:automatedML}
\end{figure}

The general workflow for training a ML model is shown in Fig. \ref{fig:automatedML}. The prepared "Data" serves as user input to train the model, while the "Target metric" checks model validity.
By combining the set of "Features" with the training "Algorithm" and "Parameters" (hyperparameters to enhance performance, model accuracy, and mitigate overfitting), we developed a model to predict Curie temperatures in chosen binary alloys.

Analysis involved combining various feature selection algorithms with training models, includes 20 different combinations ("Supplementary Materials", Table \ref{tab: algorithms}).
Among the algorithms employed are "Extremely Randomized Trees," which utilize an ensemble of multiple decision trees \cite{Geurts}, and "Gradient Boosting" \footnote{https://scikit-learn.org/stable/modules/ensemble.html}, which builds models sequentially, with each new model aimed at correcting the errors of the previous ones. We tried "LightGBM," offering enhanced scaling and faster computation \cite{Zhang}. 
We also evaluated the Elastic Net algorithm,\cite{Bai}, and the "XGBoost Regressor," known for their high accuracy in materials science tasks \cite{Heng}.

The table \ref{metrics} displays the metrics \footnote{https://learn.microsoft.com/en-us/azure/machine-learning/how-to-understand-automated-ml?view=azureml-api-2} provided by Azure ML for assessing the quality of the trained model. The primary evaluation metric is the "Normalized root mean squared error,  complemented by eleven additional metrics summarized in the "Supplementary Materials".

The prepared, analyzed, and published model is used to generate new data.

\section{Results and discussion}

This study introduces a machine-learning approach to design the Curie temperature in iron-group alloys. The central question is how to manipulate external conditions and achieve the desired magnetic transformation temperature in engineering materials. The primary factors influencing this include atomic ordering, external pressure, and the composition of the alloy.

 \subsection{Material choice,  input dataset and training model}
It is well-known that only a limited number of elements on the periodic table exhibit magnetism. All of these elements are metals, and three — iron (Fe), cobalt (Co), and nickel (Ni) — are classified as 3d transition metals within the iron group.

\begin{figure}[H]%
\includegraphics[width=1.05\textwidth]{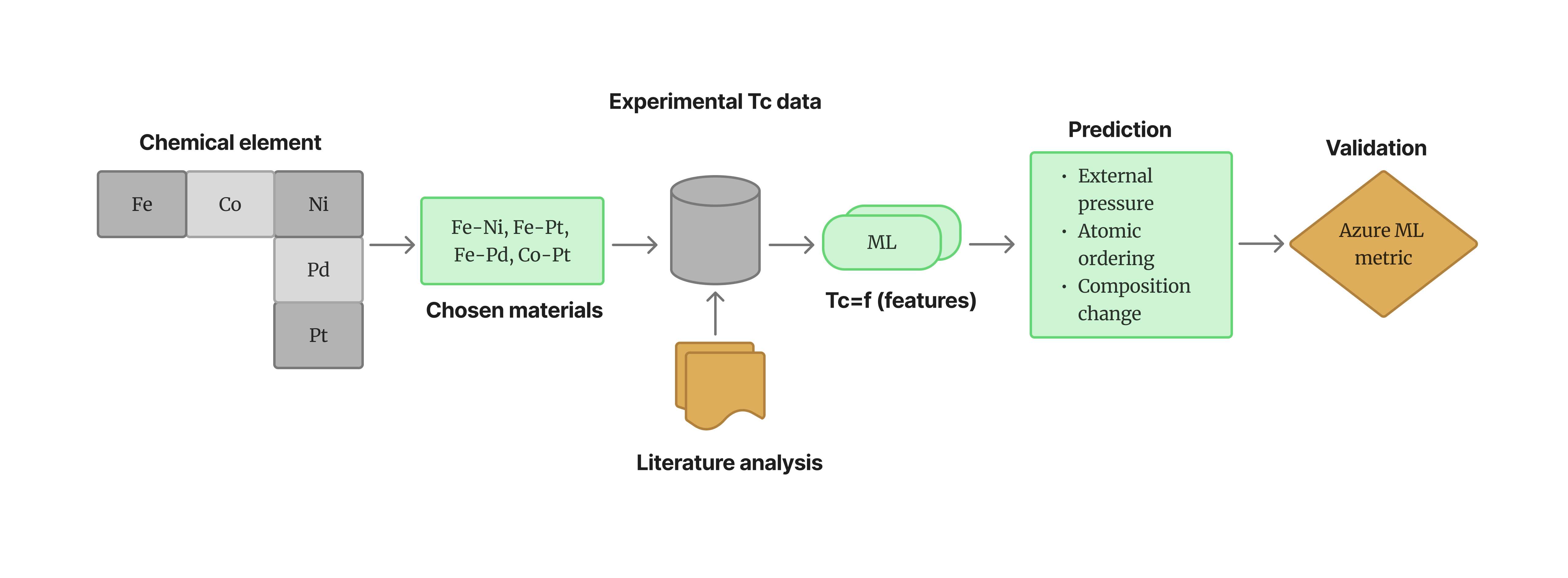}
    \caption{Data-driven strategy of alloy choise}
    \label{fig:materials}
\end{figure}

Figure \ref{fig:materials} illustrates the research strategy for selecting the objects of investigation. The data-driven search began with selecting metals, using X as the magnetic component and Y as either magnetic or non-magnetic. 
As a result, we have chosen a group of binary alloys: Fe-Pt, Fe-Pd, Co-Pt, and Fe-Ni, with the potential to expand this selection to include additional alloys. The analysis of specific features that can change the Curie temperature ($T_c$) based on published research papers allowed us to create a training dataset where $T_c$ is a function of these features: $T_c = f(features)$. The final stage of this process involves validating the model with Azure metrics (see Table \ref{metrics}).

The training dataset includes such 'features' as  external pressure $p$, concentration of X ($c_X$), and Y ($c_Y$) elements, and atomic numbers ("Atomic number X" and "Atomic number Y"), and Curie temperature $T_c$ as a 'label'
Therefore we have $Tc=f(c_x,c_y, X, Y, p, \eta)$.

To ensure a diverse dataset, experimental data for model training was collected from various publicly available sources \cite{Brannvall}, \cite{Mizoguchi}, \cite{Oomi}, \cite{Aladerah}, \cite{Wei}, \cite{Kyuji}, \cite{Wayne},  \cite{Matsushita}, \cite{Matsushita2}, \cite{Iwase}, \cite{Hayashi}, \cite{Yamamoto}, \cite{Kakeshita}, \cite{Koval}, \cite{Fallot}, \cite{Longworth}, \cite{Barmak}.
The sample of the created dataset is presented in Table \ref{tab:sample}:

\begin{table}[h]
\begin{center}
\caption{Sample of input dataset}
\label{tab:sample}
\begin{tabular}{@{}lllllll@{}}
\toprule
$c_{x}$ & $c_y$ & $Z_x$ & $Z_y$ & p & $\eta$ & $T_c$ \\
\midrule
    75    & 25 & 26 & 78 & 0 & 0 & 270\\
    65    & 35 & 26 & 28 & 0.68 & 0 & 135\\
    50    & 50 & 27 & 78 & 2.22 & 1 & 713\\
    70    & 30 & 26 & 46 & 7.8 & 1 & 400\\
\end{tabular}
\end{center}
\end{table}

Noted that the external pressure labeled "0" corresponds to the atmospheric pressure, $\eta$ = 0 means 'disordered state', and $\eta$ = 1 means atomic 'ordered state'.

The issue of overfitting is often associated with the use of small datasets \cite{Xu}. To mitigate it, we adjusted the weights for L1 and L2 regularization, varying the strength of the regularization. To address missing data in the dataset, we evaluated mean and median imputation methods. Both approaches are effective for this task.

\begin{figure}[ht]%
\includegraphics[width=1.0\textwidth]{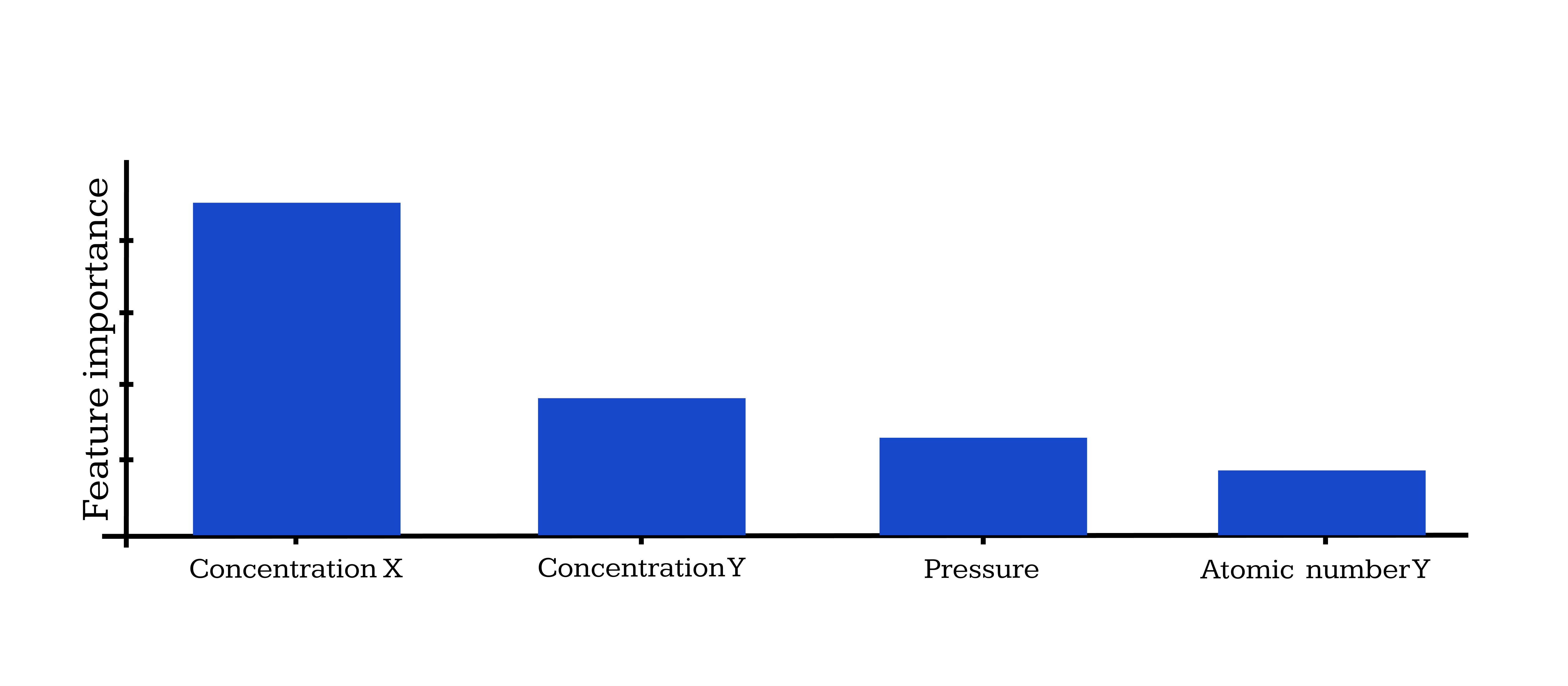}
    \caption{Feature importance for ML model predicting Curie temperature in binary alloys}
    \label{fig:features}
\end{figure}

The feature importance analyzed using Azure machine learning capabilities \footnote{https://learn.microsoft.com/en-us/azure/machine-learning/how-to-machine-learning-interpretability?view=azureml-api-2} is shown in Fig. \ref{fig:features}.

Two essential data guardrails checks, detailed in Table \ref{guardrails}, have been performed successfully.

\begin{table}[h]
\begin{center}
\caption{Dataset analysis}
\label{guardrails}%
\begin{tabular}{@{}ll@{}}
\toprule
Check      & Result \\
\midrule
    Missing feature values imputation    & Passed \\
    High cardinality feature detection    & Passed  \\
\end{tabular}
\end{center}
\end{table}

Additionally, we applied the optimization tolerance parameter \footnote{https://learn.microsoft.com/en-us/azure/machine-learning/component-reference/poisson-regression?view=azureml-api-2} as a hyperparameter. While Azure Machine Learning typically has a fixed number of training iterations, incorporating the optimization tolerance allows the model to define when it has reached an acceptable level of performance and stop iterating based on the degree of change the algorithm. This approach can improve performance by eliminating unnecessary iterations.

Various algorithms are evaluated to select the most suitable one based on Microsoft Azure's standard metrics. The best-scoring training model is prepared with the Voting Ensemble algorithm, which combines the predictions of multiple individual models to make a final prediction.
The metrics for this model are shown in Table \ref{tab: metr}.
We choose the "Normalized root mean squared error"
as a primary one for analysis, since it is a widely used performance metric to evaluate the model's accuracy in ML tasks.

\begin{table}[H]
\begin{center}
\caption{Azure ML metrics for Voting Ensemble algorithm}
\label{tab: metr}%
\begin{tabular}{@{}ll@{}}
\toprule
Metrics      & Value \\
\midrule
Explained variance     & 0.93  \\
    Mean absolute error    & 43.99\\
    Mean absolute percentage error    & 3.88  \\
    Median absolute error             & 10.19  \\
    Normalized mean absolute error    & 0.044  \\
    Normalized median absolute error  & 0.19  \\
    Normalized root mean squared error         & 0.06  \\
    Normalized root mean squared log error     & 0.53  \\
    R2 score     & 0.924  \\
    Root mean squared error         & 63.26  \\
    Root mean squared log error     & 0.17  \\
    Spearman correlation            & 0.97  \\
\end{tabular}
\end{center}
\end{table}

We investigated partitioning ratios (validation \% +testing \%) from 70\%-30\% to 80\%-20\% and 90\%-10\%. The best metrics obtained (Table \ref{tab: metr}) were for the 80\%-20\% ratio.

To assess the model's effectiveness and reliability, we must interpret key metrics.
"Explained variance" is essential for assessing a model's performance. The explained variance closer to 1.0 indicates that model predictions align closely with the actual values. In our case, the explained variance is 0.93.
Additionally, a lower "Normalized Mean Absolute Error" (NMAE) signifies that the model's predictions are more accurate compared to the actual data. Our model has a relatively low NMAE of 0.044.
The "Normalized Root Mean Squared Error" (NRMSE) also measures how closely the model's predictions match the actual values. A model is generally considered adequate if the NRMSE is 10 \% or less; in our case, it is 0.063, which is equivalent to 6.3 \%.

For our infrastructure, we used serverless parallel jobs, adjusting the number of cores and memory parameters of the virtual machine as needed. The following parameters effectively meet the requirements of our task:

\begin{table}[H]
\begin{center}
\caption{Computational details}
\begin{tabular}{@{}lc@{}}
\toprule
Memory      & Number of CPU cores \\
\midrule
    16GB RAM, 32GB storage    & 4 \\
    14GB RAM, 28GB storage    & 4  \\
    14GB RAM, 28GB storage    & 2  \\
    28GB RAM, 56GB storage    & 4  \\
\end{tabular}
\end{center}
\end{table}

\subsection{Predicted Curie temperature and Analysis}
\begin{figure}[H]%
    \includegraphics[width=1.1\textwidth]{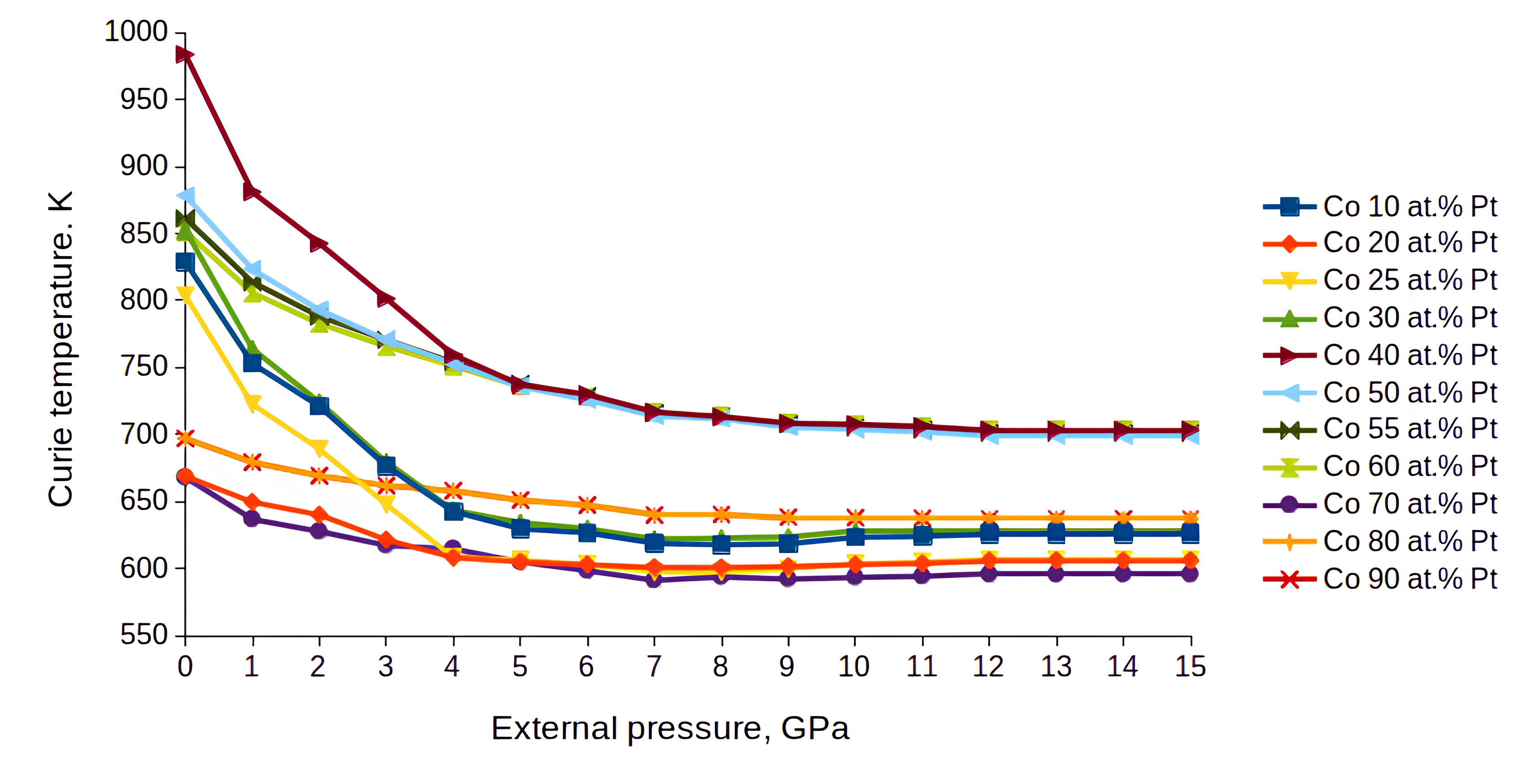}
    \caption{Pressure dependence of Curie temperature in disordered Co-Pt alloys}
    \label{fig:3}
\end{figure}

\begin{figure}[H]%
    \includegraphics[width=\textwidth]{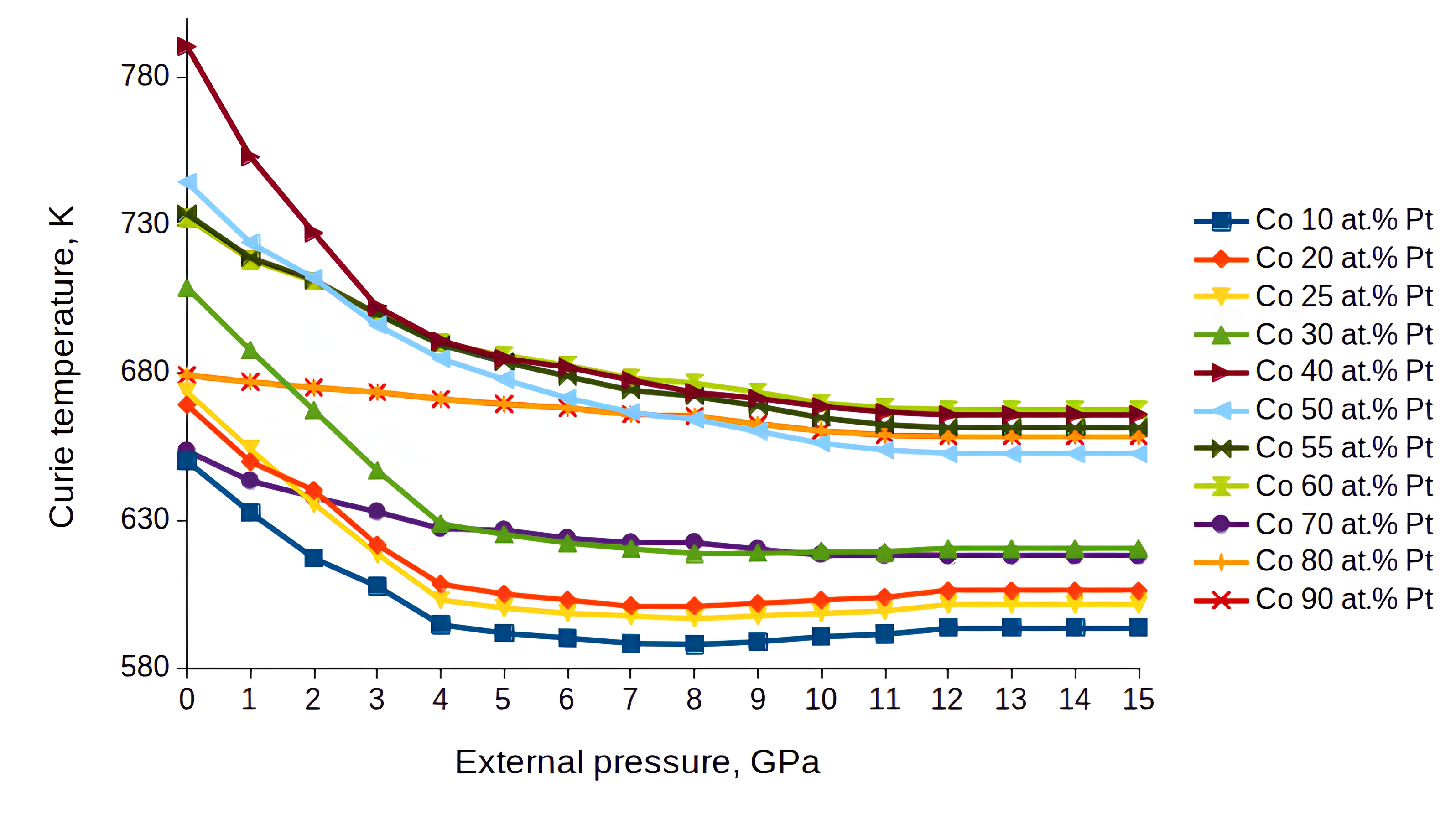}
    \caption{Pressure dependence of Curie temperature in ordered Co-Pt alloys}
    \label{fig:4}
\end{figure}

\begin{figure}[H]%
    \includegraphics[width=\textwidth]{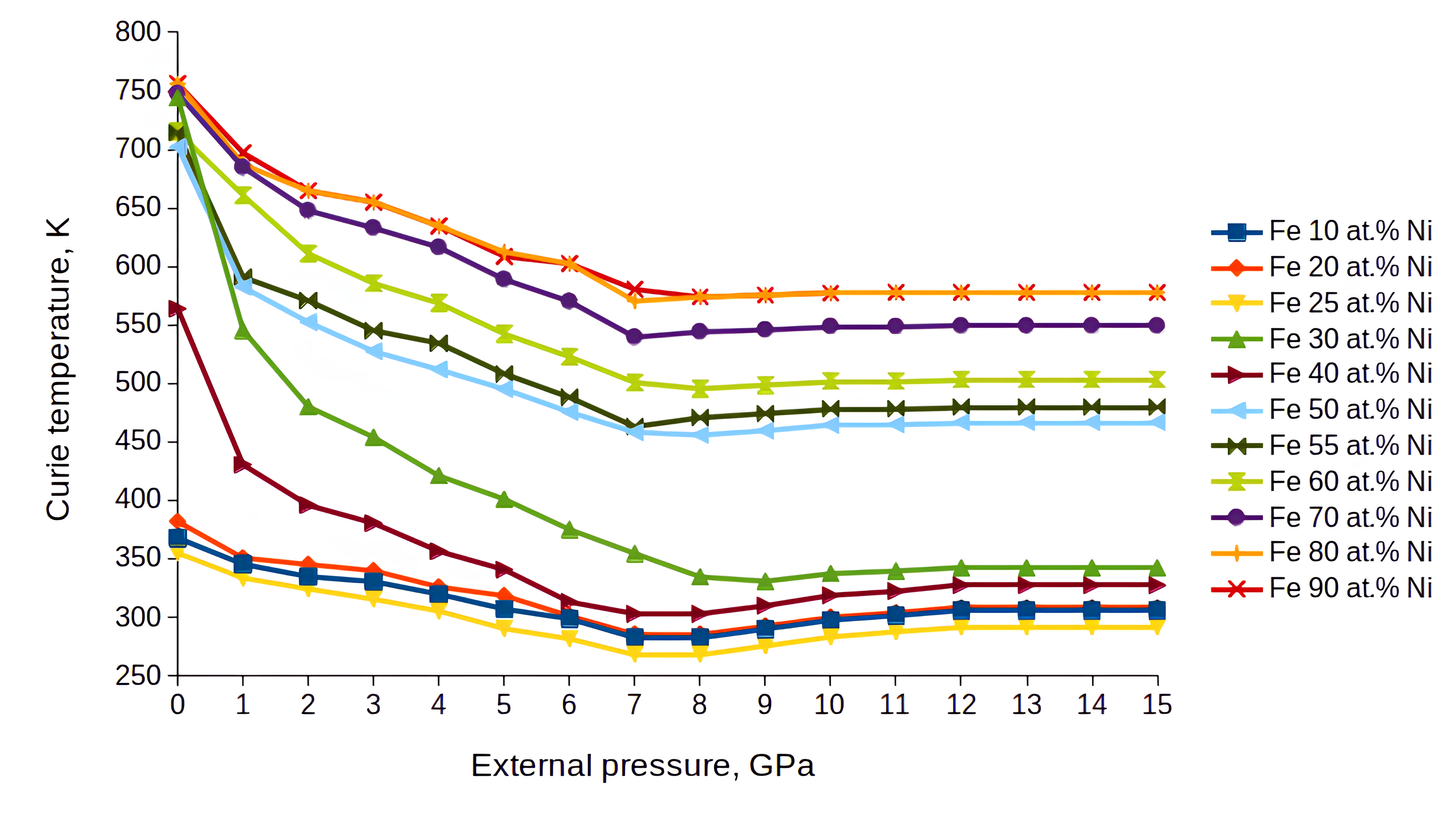}
    \caption{Pressure dependence of Curie temperature in disordered Fe-Ni alloys}
    \label{fig:5}
\end{figure}

\begin{figure}[H]%
    \includegraphics[width=\textwidth]{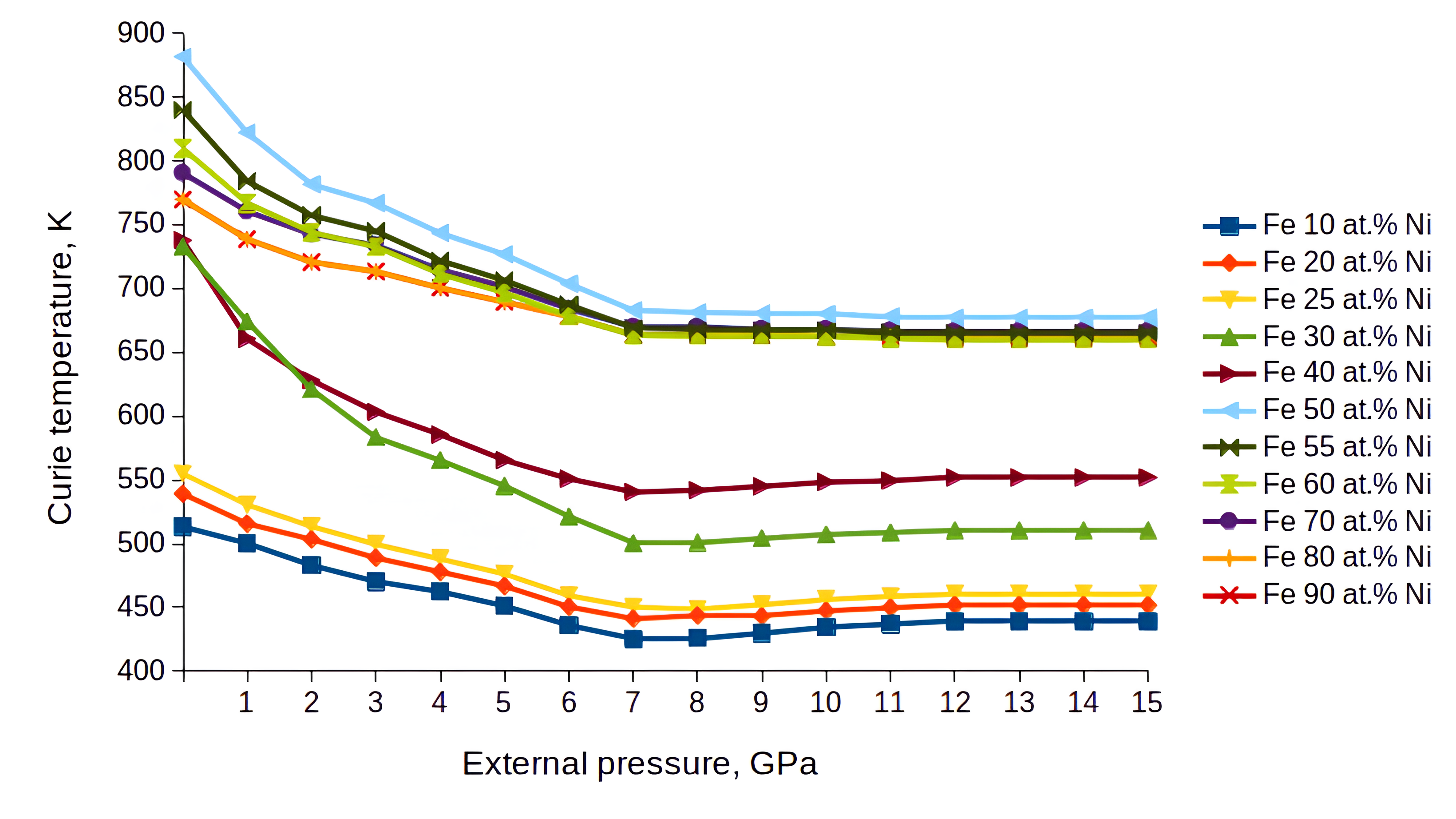}
    \caption{Pressure dependence of Curie temperature in ordered Fe-Ni alloys}
    \label{fig:6}
\end{figure}

\begin{figure}[H]%
    \includegraphics[width=\textwidth]{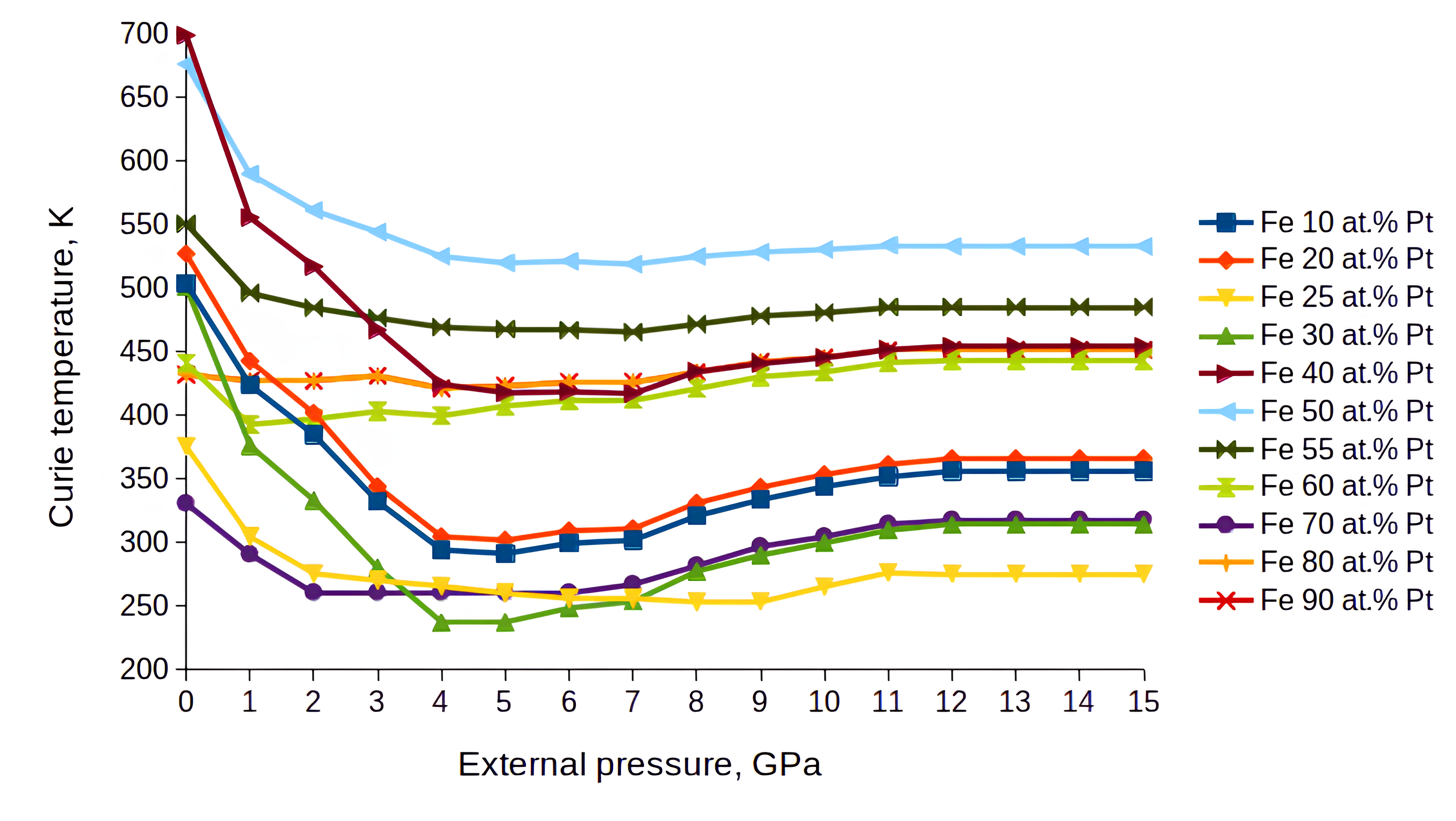}
    \caption{Pressure dependence of Curie temperature in disordered Fe-Pt alloys}
    \label{fig:7}
\end{figure}

\begin{figure}[H]%
    \includegraphics[width=\textwidth]{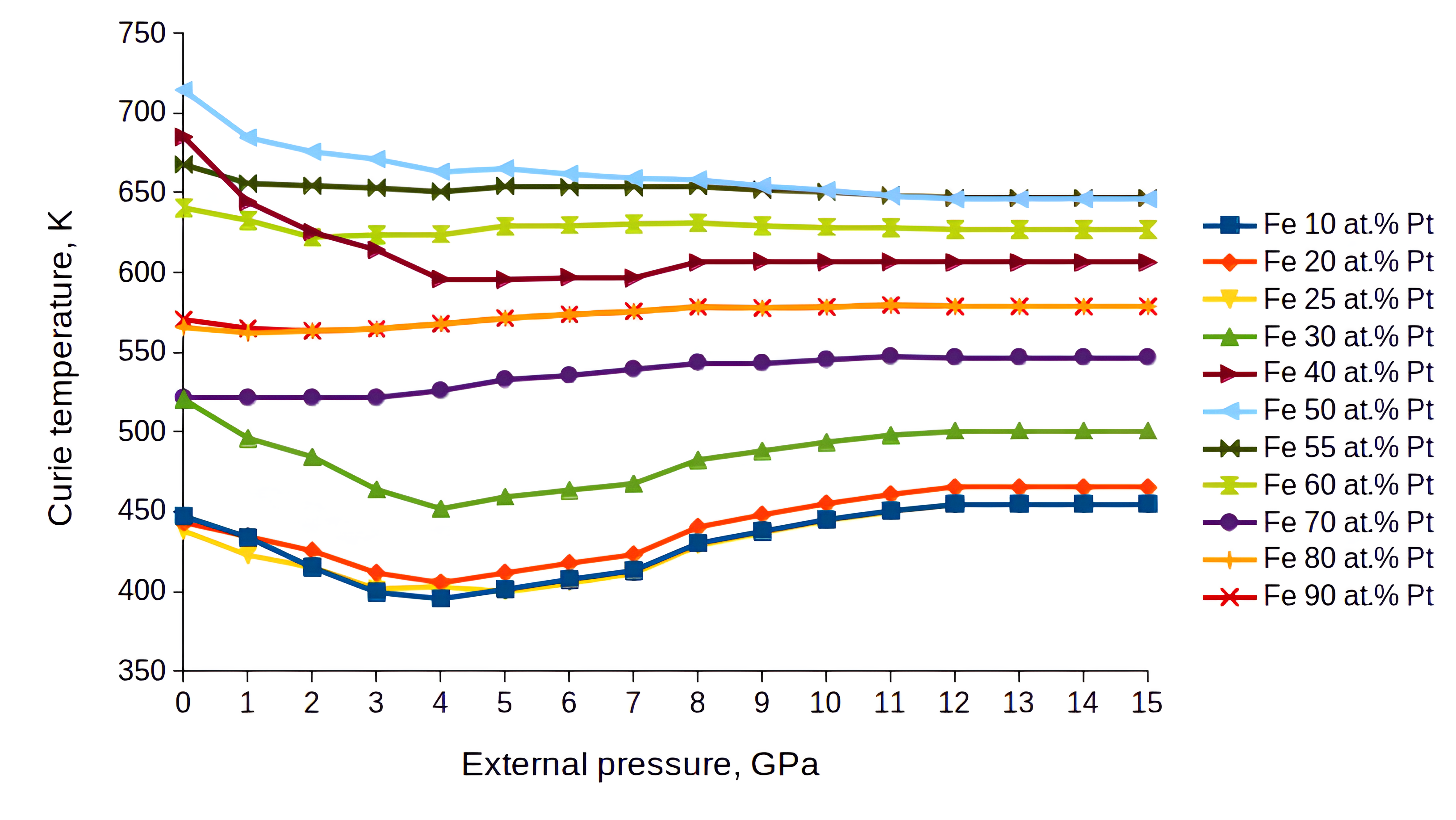}
    \caption{Pressure dependence of Curie temperature in ordered Fe-Pt alloys}
    \label{fig:8}
\end{figure}

\begin{figure}[H]%
\includegraphics[width=\textwidth]{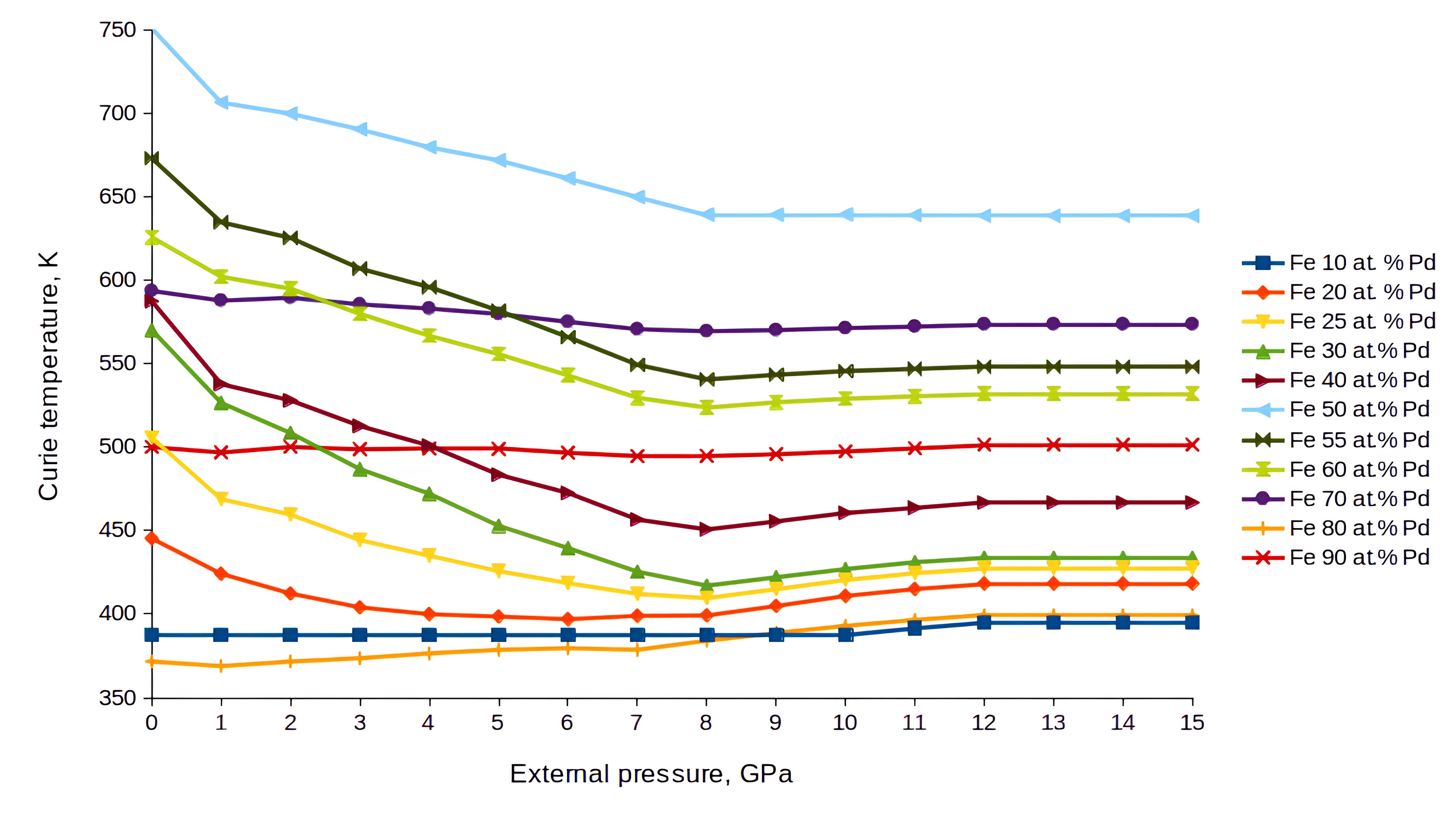}
    \caption{Pressure dependence of Curie temperature in disordered Fe-Pd alloys}
    \label{fig:9}
\end{figure}

\begin{figure}[H]%
    \includegraphics[width=1.1\textwidth]{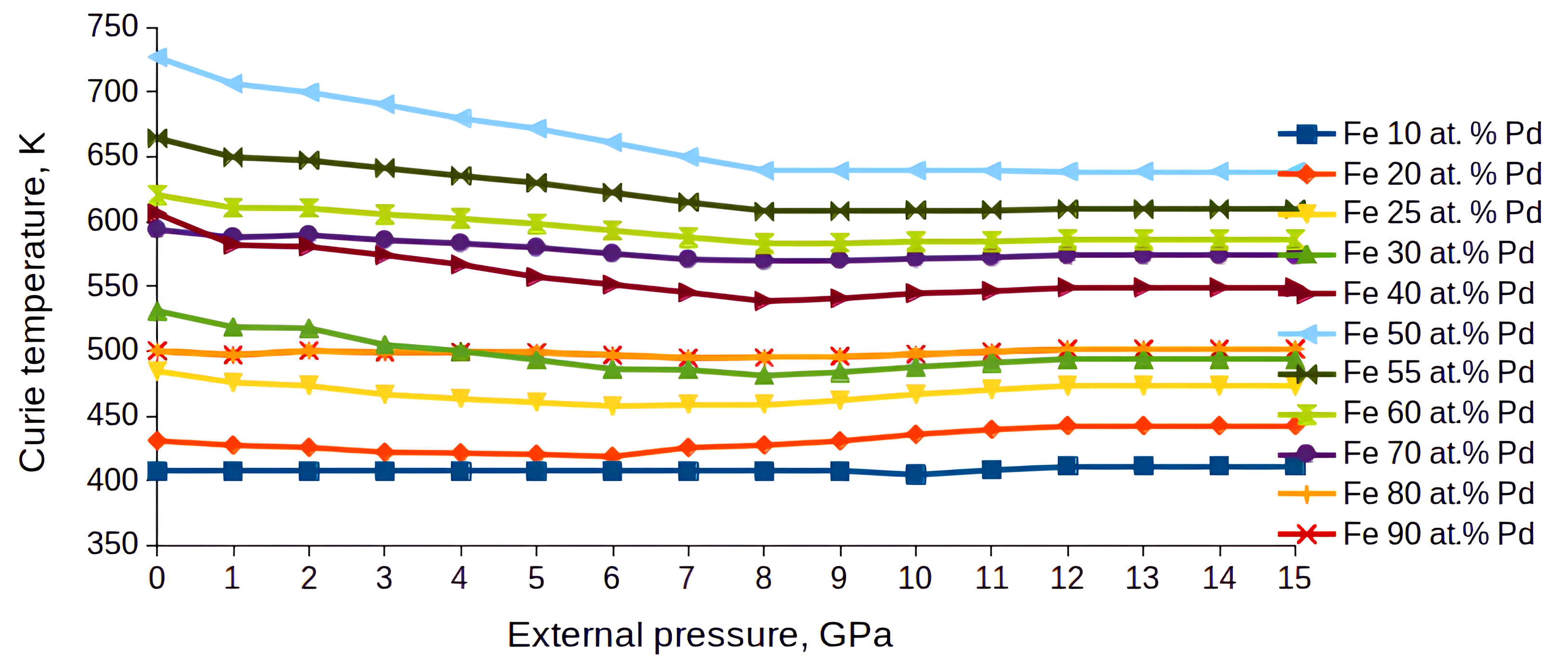}
    \caption{Pressure dependence of Curie temperature in ordered Fe-Pd alloys}
    \label{fig:10}
\end{figure}

\begin{figure}[H]%
    \includegraphics[width=\textwidth]{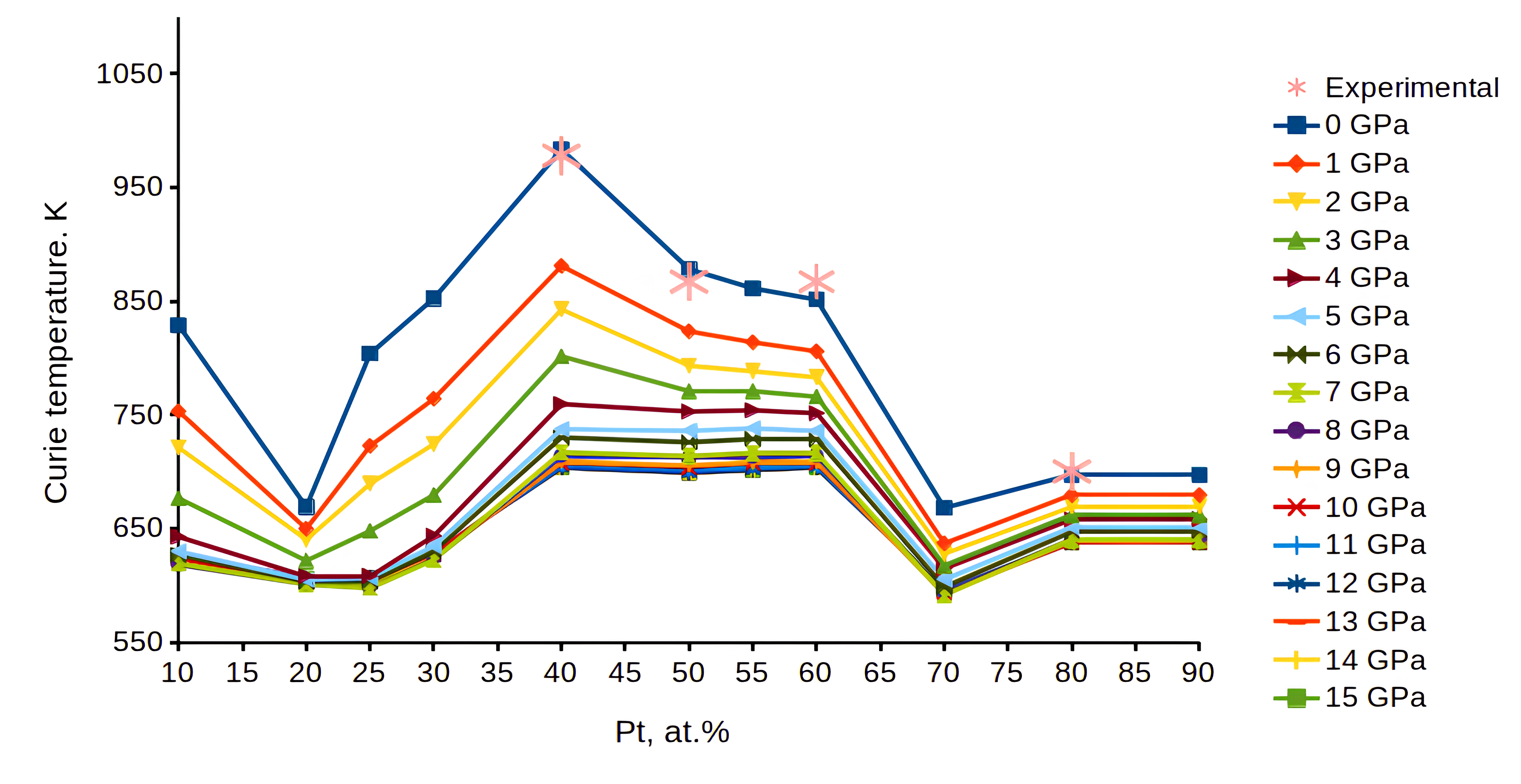}
    \caption{Concentration dependence of Curie temperature at different external pressure values in disordered Co-Pt alloys}
    \label{fig:12}
\end{figure}

\begin{figure}[H]%
    \includegraphics[width=\textwidth]{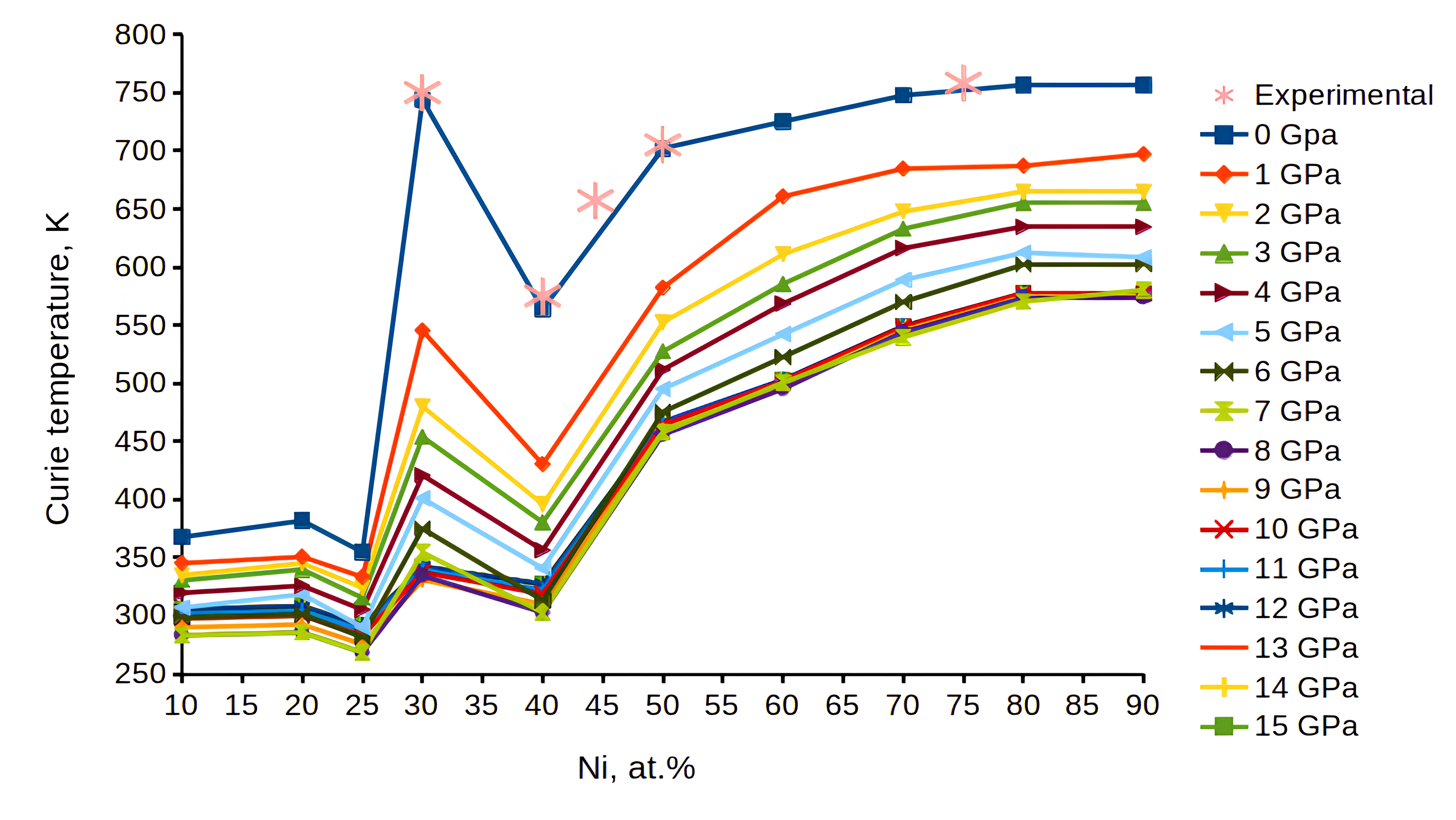}
    \caption{Concentration dependence of Curie temperature at different external pressure values in disordered Fe-Ni alloys}
    \label{fig:13}
\end{figure}

\begin{figure}[H]%
    \includegraphics[width=\textwidth]{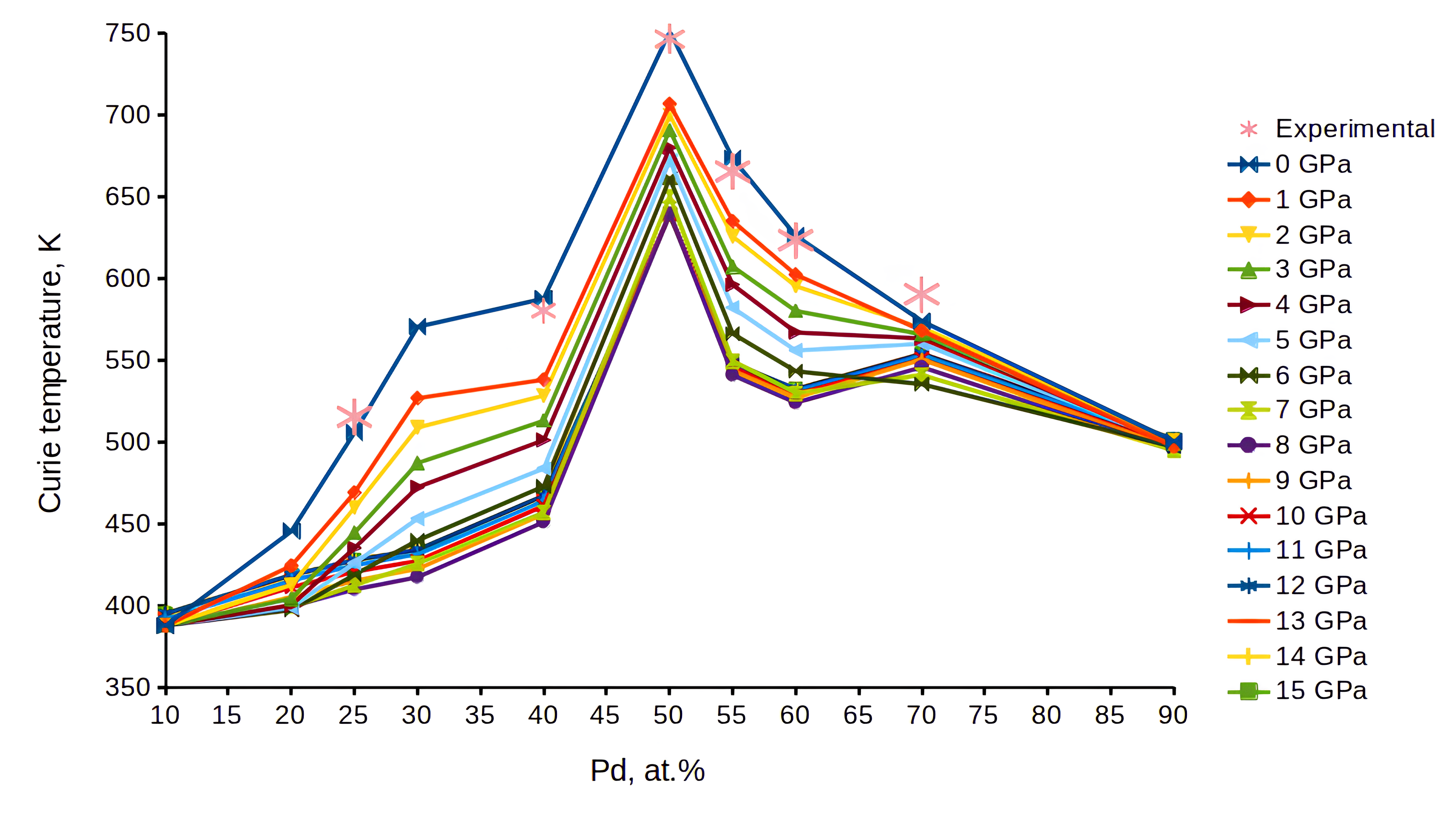}
    \caption{Concentration dependence of Curie temperature at different external pressure values in disordered Fe-Pd alloys}
    \label{fig:14}
\end{figure}

\begin{figure}[H]%
    \includegraphics[width=\textwidth]{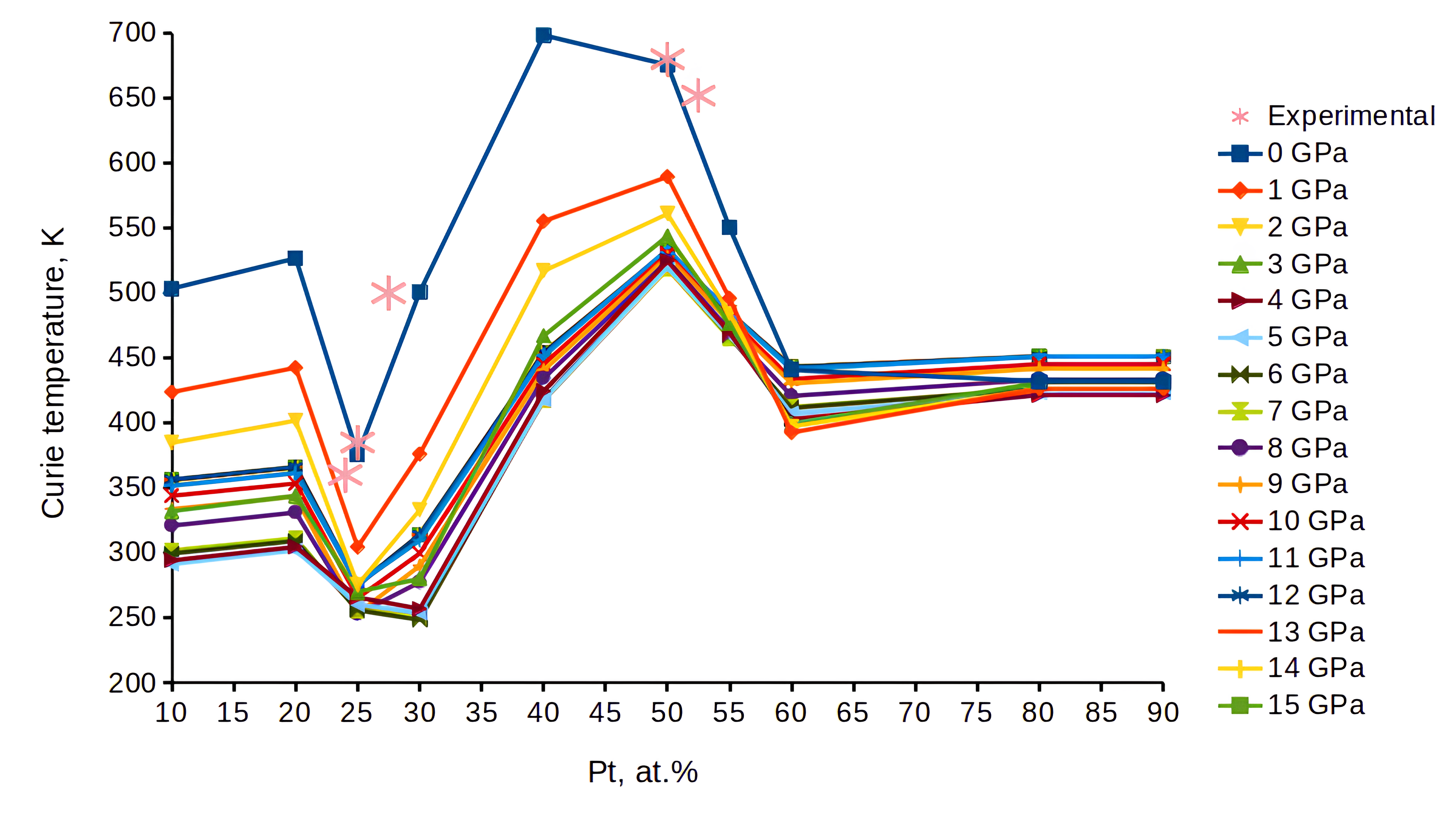}
    \caption{Concentration dependence of Curie temperature at different external pressure values in disordered Fe-Pt alloys}
    \label{fig:15}
\end{figure}

\begin{figure}[H]%
    \includegraphics[width=\textwidth]{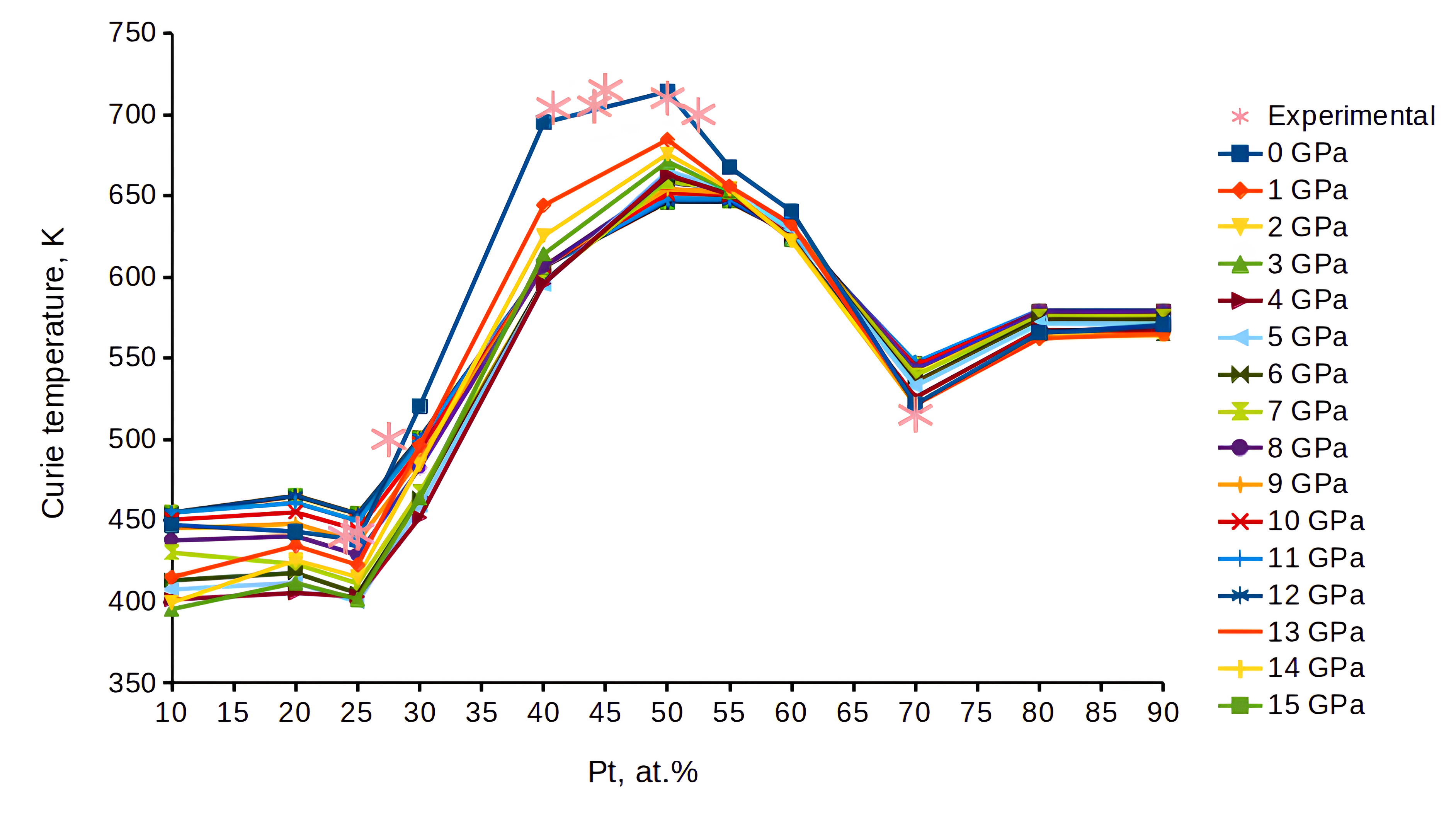}
    \caption{Concentration dependence of Curie temperature at different external pressure values in ordered Fe-Pt alloys}
    \label{fig:16}
\end{figure}

Figures \ref{fig:3}-\ref{fig:16} illustrate the predicted by ML Curie temperatures for Co-Pt, Fe-Pt, Fe-Pd, and Fe-Ni alloys across various compositions, both in their ordered and disordered states, under different external pressures.

To establish the range of external pressures for the machine learning predictions, we examined the critical pressure at which ferromagnetism ceases to exist. According to Aladerah \cite{Aladerah}, critical external pressure varies from 50 to 164 GPa, so taking the range from atmospheric pressure to 15 GPa for investigation is physically acceptable.

We experimented with various alloy concentrations across the entire range, having X-(10, 20, 25, 30, 40, 50, 55, 60, 70, 80, 90) at.\% Y alloys where X represents Fe or Co and Y = Ni, Pt, or Pd.

As for atomic ordering influence, the most significant changes among all are observed in Co-Pt alloys (Figs. \ref{fig:3} and \ref{fig:4}). The highest reached Curie temperature (about 983 K) is seen for Co-40 at.\%Pt in the disordered state at lower levels of external pressure, decreasing it from 983 to 801 K in the range of pressure 0-3 GPa. Disordered Co-Pt alloys (Fig. \ref{fig:3}) exhibit a significantly higher Curie temperature than their ordered counterparts (Fig. \ref{fig:4}). In contrast, investigations for Fe-Ni alloys (Fig. \ref{fig:5}, \ref{fig:6}) and Fe-Pt alloys (Fig. \ref{fig:7}, \ref{fig:8}) show that atomic ordering increases their Curie temperature.

The external pressure lowers the Curie temperature for all binary alloys studied in this research. 
That is consistent with existing experimental research. Notably, the authors of the paper \cite{Aladerah} demonstrated a reduction in the Curie temperature as pressure increased in Fe-Ni alloys. Additionally, the paper \cite{Wayne} provides experimental evidence indicating a decrease in the Curie temperature for Fe alloyed with Ni, Pd, and Pt at concentrations of (25-40) atomic percent. These studies support and confirm our model results.

With varying concentrations and ordering states, we observe a tendency for the Curie temperature to approach an asymptotic value at specific levels of external pressure. This asymptotic pressure level is lower for Co-Pt and Fe-Ni alloys (Fig. \ref{fig:3}-\ref{fig:6}). It is approximately 7-8 GPa, while it is higher for Fe-Pd and Fe-Pt alloys, ranging between 9 and 10 GPa (Fig. \ref{fig:7} - \ref{fig:10}).

The relationship between concentration and Curie temperature is non-linear, exhibiting a distinct maximum that varies depending on the alloy. Notably, the maximum Curie temperature in Co-Pt alloys occurs at 40 atomic percent (at.\% Pt) (see Fig. \ref{fig:12}), while for Fe-Pt alloys, it is also observed at 40 at.\% Pt (see Fig. \ref{fig:15}). In the case of Fe-Ni alloys, the maximum Curie temperature is reached at 30 at.\% Ni. For Fe-Pd alloys, the maximum is observed at 50 at.\% Pd (see Fig. \ref{fig:14}). In contrast, the Curie temperature remains nearly constant for platinum-rich, palladium-rich, and nickel-rich alloys with concentrations ranging from 70 to 90 at.\% Pt and Ni. This behavior has been documented in Co-Pt (see Fig. \ref{fig:12}), Fe-Ni (see Fig. \ref{fig:13}), Fe-Pt (see Fig. \ref{fig:14}), and Fe-Pd (see Fig. \ref{fig:15}) alloys. The predicted values of Curie temperatures have been compared with experimental results, demonstrating good correspondence.

We demonstrated how the combination of composition changes and atomic ordering affects the Curie temperature, using Fe-Pt as an example. In the disordered state (see Fig. \ref{fig:14}), the Curie temperature exhibited two minima: one around the composition of $Fe_3Pt$ (approximately 25 at.\% Pt) and another at the stoichiometry of $FePt_3$ (about 70 at.\% Pt). The shape of the curve depicting composition dependence changes significantly when atomic ordering is present (see Fig. \ref{fig:15}). While it still shows two minima, the drastic reduction in the Curie temperature at 25\% Pt is not observed, and the changes in the Curie temperature occur more smoothly.

The phase stability for the alloy as a function of external pressure ($p$), alloy concentration ($c$), and degree of atomic order ($\eta$) can be represented by the thermodynamic potential $F(p, c, \eta)$. By applying the Legendre transformation \cite{Kubo}, we can convert the function to enthalpy $H$, changing its variables from (p, c, $\eta$) to (p, T, N): $F (p, c, \eta) \rightarrow H(p, T, N)$.

Enthalpy change can be represented by \cite{Kubo}:
\begin{align}
  \Delta H(p, T, N) = T \Delta S + V \Delta p + \mu \Delta N  \rightarrow  \\
  \bigg(\Delta S_{conf}(p, N) + \Delta S_{vib}(p, T)\bigg)T + V \Delta p + \mu \Delta N
  \label{entalpy}
\end{align}
with $S_{conf}$, $S_{vib}$ as the configurational and vibrational entropy change.

Atomic ordering and alloy composition vary the configuration part of $H(p, T, N)$ -  $S_{conf}(p, N)$, as it reflects randomness in the arrangement of atoms. On the other hand, it is indeed related to the electron density of states (DOS) $g(\epsilon)$, which quantifies the number of available microstates at a given energy level.
    \begin{equation}
		\Delta S_{conf}(p, N)  \approx 3k_B(g(\epsilon) - g^{0}(\epsilon))
    \end{equation}
    \label{S_conf}
\newline
Here $g^{0}(\epsilon)$ is the initial electron density of states in the alloy; $g(\epsilon)$ is the electron density of states under external influences, and $k_B$ is the Boltzmann constant.

The vibrational entropy also depends on external pressure \cite{Fultz}, and temperature, and they control the phonon density of states, $g(\omega)$(\cite{Tsuyama}, \cite{Yu}:
\begin{equation}
    \Delta S_{\mathrm{vib}} (p, \eta)=3 k_{\mathrm{B}} \int_{0}^{\infty}
    \big(g(\omega) - g^{0}(\omega) \big)
    ln(\omega)\mathrm{d} \varepsilon d\omega
    \label{vibration}
\end{equation}
\newline

Therefore, the enthalpy $H(p, T, N)$, which describes macro state of the alloy, can be expressed in terms of the micro parameters of the system - the electron DOS $g(\epsilon)$ and phonon DOS $g(\omega)$:
$\Delta H (p, T, N) \rightarrow H(g(\epsilon),g(\omega)\big)$

The decreasing Curie temperature with pressure can be easily understood through the entropy change. When applied, external pressure provides the atoms with additional kinetic energy. As the movement of atoms increases due to vibrations, the magnetic order is disrupted at lower temperatures. Consequently, as pressure increases, more kinetic energy and vibrational entropy are introduced, lowering the Curie temperature. This behavior explains the general relationship between pressure and the Curie temperature across different systems.
The asymptotic value of $T_C$ can be attributed to the saturation of vibrational energy and entropy within the system at a certain external pressure.

As experimentally seen \cite{Watanabe}, in Fe-Pd alloys, the density of states at the Fermi level is decreased with increased atomic ordering. This tendency is also observed for Fe-Ni alloys. 
In contrast, the DOS is increased with atomic ordering in Fe-Pt \cite{Tsuyama} and Co-Pt alloys.
The decrease in the Curie temperature with atomic ordering is correlated with a reduction in the density of states for Fe-Pd and Co-Pt alloys. An increase in the DOS corresponds to a rise in the Curie temperature for Fe-Pt and Fe-Ni alloys.

 As shown, external pressure, atomic ordering, and changes in alloy composition are important factors in designing the Curie temperature for binary alloys. Pressure consistently lowers the Curie temperature for all the alloys examined: Co-Pt, Fe-Pt, Fe-Pd, and Fe-Ni. 
 Atomic ordering can either decrease the Curie temperature (as seen with Co-Pt and Fe-Pd) or increase it (as observed in Fe-Pt and Fe-Ni). In the latter case, external pressure and atomic ordering are competing factors. Pressure's effect dominates at 5-7 GPa for Fe-Pt alloys and about 7 GPa for Fe-Ni alloys. 

Table \ref{tab:control} shows the analysis of control factors that affect the Curie temperature. For instance, to achieve the highest possible Curie temperature for Co-Pt alloys, the composition should be close to Co-40 atomic percent (at.\%) Pt, maintained under atmospheric pressure and in a disordered state, since atomic ordering and high external pressure lower the Curie temperature. 
Conversely, to obtain the lowest Curie temperature for Fe-Pt alloys, the ideal composition would be about Fe-70 at.\% Pt, also in a disordered state, while applying external pressure of 4 to 6 GPa. This demonstrates that we can design the Curie temperature by controlling three key parameters.

\begin{table}[H]
    \caption{Control factors to design Curie temperature ($T_c$) in binary alloy}
    \centering
    \begin{tabular}{l|l|l|p{70mm}}
        \hline
        \textbf{Alloy} & \textbf{External pressure} & \textbf{Atomic ordering} & \textbf{Alloy composition change \newline }\textit{(in disordered state under normal pressure)}\\
        \hline
            Co-Pt & Decreases $T_c$ & Decreases $T_c$ & * 10-20 at.\% Pt slightly increases $T_c$; \newline * 20-40 at.\% Pt significantly increases $T_c$; \newline * 40 at.\% Pt - the highest $T_c$;
            \newline * 40-60 at.\% Pt - slightly decreases $T_c$; \newline * 60-70 at.\% Pt significantly $T_c$; \newline * 70-90 at.\% Pt keeps asymptotic $T_c$. \newline \\
       \hline
            Fe-Pd & Decreases $T_c$ & Decreases $T_c$ & * 10-30 at.\% Pd slightly increases $T_c$; \newline * 30-50 at.\% Pd significantly increases $T_c$; \newline * 50 at.\% Pd - the highest $T_c$;
            \newline * 50-70 at.\%  Pd significantly decreases $T_c$; \newline * 70-90 at.\% Pd slightly reduces $T_c$. \newline \\
        \hline
            Fe-Ni & Decreases $T_c$ & Increases $T_c$ & * 10-25 at.\% Ni slightly increases $T_c$; \newline * 25 at.\% Ni - first minimum of  $T_c$; \newline * 25-30 at.\% Ni significantly increases $T_c$; \newline * 30 at.\% Ni - maximum of  $T_c$ \newline * 30-40 at.\% Ni - decrease to the second min; \newline -40 at.\% Ni - the second minimum of $T_c$; \newline * 40-65 at.\% Ni increases $T_c$;
            \newline * 65-90 at.\% Ni $T_c$ - keeps asymptotic value.  \newline\\
        \hline
            Fe-Pt & Decreases $T_c$ & Increases $T_c$ & * 10-20 at.\% Pt slightly increases $T_c$; 
            \newline * 20-25 at.\% Pt fast decrease to minimum;
            \newline * 25-40 at.\% Pt significantly increases $T_c$; \newline * 40 at.\% Pt - the highest $T_c$;
            \newline * 40-60 at.\% Pt reduces $T_c$ to the minimum; \newline * 60-90 at.\% Pt asymptotic value. \newline \\
        \hline
        \end{tabular}
    \label{tab:control}
\end{table}

\newpage
\section{Summary}

This study explores the use of machine learning algorithms to predict the Curie temperature of binary alloys, specifically Fe-Pt, Fe-Pd, Fe-Ni, and Co-Pt. It focuses on several key factors:
\newline
- variations in element concentration;
\newline
- the influence of external pressure;
\newline
- the degree of atomic ordering.

The features utilized in this study include specific characteristics of the alloys. These characteristics consist of the concentration of elements in the binary alloy ($c$), external pressure ($p$), the degree of atomic order ($\eta$), and the atomic numbers of the elements in the alloy, all of which correlate with the Curie temperature. To extend the application of this model, the feature set can be modified to predict other magnetic properties of iron group binary systems by replacing the Curie temperature with other relevant parameters, such as coercive force and others.

Several algorithms were analyzed to train the regression model, including Random Forest, Extreme Random Trees, Decision Tree, Gradient Boosting, ElasticNet, XGBoost Regressor, LightGBM, and Voting Ensemble. Among these, the Voting Ensemble algorithm was found to be the most effective based on the performance metrics evaluated (see Table \ref{tab: metr}).

The study shows that external pressure decreases the Curie temperature in all examined binary alloys. Interestingly, atomic ordering has varying effects: it increases the Curie temperature for Fe-Pt and Fe-Ni alloys, while it decreases it for Fe-Pd and Co-Pt alloys. These results align with experimental observations regarding the Curie temperature.
The change in concentration leads to a non-linear effect. Initially, with a low concentration of approximately 10-20 atomic percent (at.\%) of Pt, Pd, or Ni in iron (Fe) or cobalt (Co), the Curie temperature increases. As the concentration continues to rise, the Curie temperature reaches a peak in the middle of the concentration range, before declining. At higher concentrations, around 70-90 at.\% (in some cases starting from 60 at.\%), the Curie temperature shows only slight concentration dependence.

Consequently, the combined effects of atomic ordering, concentration variation, and external pressure can be utilized to design materials with specific Curie temperatures. The proposed features, computational methods, and feature analysis algorithms can be applied to a variety of materials, including Ni-Pt, Ni-Pd, Co-Pd alloys, as well as ternary alloys and bulk materials. This set of features can also be adapted for nanoparticles by taking into account their shape and size.

\section{Conflict of interest}
The authors declare no conflict of interest.

\section{Acknowledgment}
This paper would not have been possible without the exceptional support of Alla Lysak.

\section{Data Availability Statement}
Data is available per reasonable request.

\section{Funding details}
No special funding was provided for this work.

\section{Disclosure statement}
The authors report there are no competing interests to declare.

\bibliography{mybibfile}

\section*{Supplementary materials}

\begin{table}[H]
\begin{center}
\caption{Data preparation and training algorithm pairs}
\label{tab: algorithms}%
\begin{tabular}{@{}lll@{}}
\toprule
Feature selection algorithm  & Model training algorithm \\
\midrule
Standard Scaler Wrapper& Extreme Random Trees \\
MaxAbsScaler& Extreme Random Trees \\
Sparse Normalizer& Extreme Random Trees \\
Standard Scaler Wrapper& Random Forest \\
MaxAbsScaler& Random Forest \\
Standard Scaler Wrapper& Gradient Boosting \\
TruncatedSVD Wrapper& Extreme Random Trees \\
Standard Scaler Wrapper& ElasticNet \\
MaxAbsScaler& ElasticNet \\
Standard Scaler Wrapper& Decision Tree \\
Standard Scaler Wrapper& LightGBM \\
MaxAbsScaler& LightGBM \\
MaxAbsScaler& Decision Tree \\
MaxAbsScaler& Gradient Boosting \\
Sparse Normalizer& Random Forest \\
Standard Scaler Wrapper& XGBoost Regressor \\
Sparse Normalizer & Voting Ensemble \\
Standard Scaler Wrapper & Voting Ensemble \\
TruncatedSVD Wrapper & Voting Ensemble \\
MaxAbsScaler & Voting Ensemble \\
\end{tabular}
\end{center}
\end{table}

\begin{table}[H]
\begin{center}
\caption{Azure ML metrics and short explanation}
\label{metrics}%
\begin{tabular}{@{}ll@{}}
\toprule
Metrics    \\
\midrule
\textbf{Explained variance}: percent decrease in variance of the original data to the variance of the errors   \\
\textbf{Mean absolute error}: difference between the target and the prediction    \\
\textbf{Mean absolute percentage error}: average difference between a predicted and actual values   \\
\textbf{Median absolute error}: median difference between the observations and model output   \\
\textbf{Normalized mean absolute error}: evaluates the accuracy of models    \\
\textbf{Normalized median absolute error}: difference between a dataset predicted and actual values \\
\textbf{Normalized root mean squared error} : evaluates model performance across different datasets \\
\textbf{Normalized root mean squared log error}: compares the normalized predicted to observed values   \\
\textbf{R2 score}: shows variability in the data    \\
\textbf{Root mean squared error}: average difference between the projected and actual values   \\
\textbf{Root mean squared log error}: average difference between the logarithm of the predicted and actual values     \\
\textbf{Spearman correlation}: monotonicity for dataset         \\
\end{tabular}
\end{center}
\end{table}

\end{document}